\documentclass{llncs}

\usepackage{amsmath,amssymb}
\usepackage{graphicx}
\usepackage{wrapfig}
\usepackage{caption}
\usepackage{subcaption}
\usepackage{verbatim}
\usepackage{paralist}
\usepackage{cancel}
\usepackage{xspace}
\usepackage{cleveref}
\usepackage{mathtools}
\usepackage{algorithm}
\usepackage{algorithmicx,algpseudocode}
\usepackage{makecell}
\usepackage[dvipsnames]{xcolor}
\usepackage{cite}
\usepackage{booktabs}
\usepackage{placeins}
\usepackage{tikz,ifthen,pgfplots,forest}
\definecolor{navy}{RGB}{0,0,128}
\definecolor{dodgerblue}{RGB}{30,144,255}
\usepackage{marginnote}

\usepackage{xspace}
\newcommand{\relu}{\text{ReLU}\xspace{}}

\usepackage{todonotes}

\usepackage{marvosym}
\newcommand{\mysubsection}[1]{\medskip\noindent\textbf{#1}}

\title{Learning to Split: A Reinforcement-Learning-Guided Splitting Heuristic for Neural Network Verification}
\author{Maya Swisa \and Guy Katz\textsuperscript{(\Letter)}}
\institute{The Hebrew University of Jerusalem, Jerusalem, Israel \\ \email{ \{maya.swisa, g.katz\}@mail.huji.ac.il}}

\begin{document} 
\maketitle

\begin{abstract}
  Modern neural network verifiers often encode neural network
  verification as constraint satisfaction problems.  When dealing with
  standard piecewise-linear activation functions, such as ReLUs,
  verifiers typically employ branching heuristics that break a complex
  constraint satisfaction problem into multiple, simpler problems. The
  verifier's performance depends heavily on the order in which this
  branching is performed: a poor selection may give rise to
  exponentially many sub-problems, hampering scalability. Here, we
  focus on the setting in which many related verification queries must
  be solved for the same neural network.  The core idea is to use past
  experience to make good branching decisions, expediting
  verification.  We present a reinforcement-learning-based branching
  heuristic that achieves this, by applying Deep Q-learning from
  Demonstrations (DQfD). Our experimental evaluation demonstrates a
  substantial reduction in average verification time and in the
  average number of iterations required, compared to modern splitting
  heuristics.  These results highlight the great potential of
  reinforcement learning in the context of neural network
  verification.
\end{abstract}

\section{Introduction}\label{sec:Introduction}
Deep neural networks (DNNs) have emerged as highly powerful tools,
which are increasingly being integrated into complex systems in the
fields of computer vision~\cite{HeZuReSu16}, natural language
processing~\cite{Op22}, and aerospace~\cite{KaLeReYe24}.  However,
DNNs are also prone to mistakes: for example, they will often fail
when presented with mildly perturbed (``adversarial'')
inputs~\cite{szegedy2013intriguing,GoShSz15,ChWuGuLiJh17}.  The
growing reliance on DNNs, combined with their susceptibility to error,
highlights the need for formal guarantees regarding their behavior;
and indeed, great strides have been made on DNN verification in the
past decade to try and meet this goal.

Despite these advancements, a substantial disparity remains between
the scale of DNNs amenable to current verification techniques and the
complexity of those required for real-world
applications~\cite{BrBaJoWu24}. This is primarily due to the fact that DNN
verification with piecewise-linear activations is
NP-complete~\cite{KaBaDiJuKo17}, indicating that the computational
cost can grow exponentially with the size of the network. This
intractability necessitates the further development of sophisticated
and efficient verification techniques.

Many modern verifiers operate by translating the verification problem
into the Satisfiability Modulo
Theories (SMT) setting (alongside other techniques, such as symbolic
bound propagation~\cite{WaZhXuLiJaHsKo21,BrBaJoWu24}). This approach
provides sound and complete guarantees for a DNN's correctness, or a
counterexample demonstrating the undesirable behavior. While these
methods guarantee a definitive answer, they rely on \emph{case
  splitting} --- the translation of piecewise-linear constraints to
disjunctions of linear constraints --- to handle the piecewise-linear
activation functions of the DNN. This process transforms a complex
problem into a long sequence of simpler problems, and is known to
cause scalability issues~\cite{KaBaDiJuKo17}. It is known that the
selection of neurons on which a solver splits can have a tremendous
effect on the size of the search space that is eventually explored;
and so multiple splitting heuristics have been proposed with the goal
of minimizing this search space (e.g.,
Pseudo-Impact~\cite{WuZeKaBa22},
polarity~\cite{WuOzZeJuIrGoFoKaPaBa20}, and
BaBSR~\cite{BuLuTuToKoKu20}).

A fundamental limitation of existing approaches is that a single,
static heuristic is typically chosen and applied throughout the entire
verification process. This method overlooks three key points. First,
it is very difficult to know, a priori, which heuristic will be the
most useful for the problem at hand; second, it may be useful to
change heuristics during the verification procedure, as different
parts of the search space are traversed; and third, in many
applications, a verification query is actually not stand-alone, but
part of a set of related queries --- and in that case, it can be
useful to transfer knowledge from one run of the verifier to the next,
potentially reducing the overall verification time.

To address these limitations, we propose a novel approach: we
use Double DQN with demonstrations (DQfD) to learn a
state-conditioned, adaptive splitting policy. Rather than fixing a
single heuristic a priori, we train an agent that selects the next
split according to an estimated action value at the current verifier
state. This enables state-dependent branching decisions, allows the
preferred split to change as the search evolves, and exploits
experience from related verification queries through demonstrations.

To evaluate our approach, we implemented it on top of the Marabou
solver~\cite{KaHuIbJuLaLiShThWuZeDiKoBa19}, and trained and evaluated
the learned policy on the ACAS Xu benchmark
family~\cite{KaBaDiJuKo17}.  We considered both the standard ACAS Xu
safety properties~\cite{JuKoOw19} and local robustness queries,
requiring the predicted advisory to remain unchanged within an
$\ell_\infty$ ball around a given input.  Compared with Marabou's
built-in splitting heuristics, our learned policy solved more
instances and reduced average verification time, with  
improvements most evident on harder instances. Across the evaluated settings, the
average time reduction ranged from 5.88\% to 56.20\%, highlighting the
potential of learned branching for DNN verification.

The rest of the paper is organized as follows.
\Cref{sec:preliminaries} provides background on deep neural networks, their
verification, and reinforcement learning.  In \Cref{sec:methodology},
we present our methodology for learning adaptive splitting policies.
\Cref{sec:results} reports on implementation details and experimental
evaluation on safety and robustness benchmarks, followed by
Section~\ref{sec:related_work} where we discuss related work.
Finally, Section~\ref{sec:conclusion_and_future_work} concludes and
outlines directions for future work.

\section{Preliminaries}\label{sec:preliminaries}
\subsection{Neural Networks and their Verification}
Deep neural networks (DNNs) are composed of layers of interconnected
neurons. The output of neuron \(j\) in layer \(l\) is
\(a_j^{(l)} = \sigma(W_j^{(l)}a^{(l-1)} + b_j^{(l)})\), where
\(\sigma\) is the activation function. We focus on the ReLU activation
function, \(\relu(x)=\max(0,x)\). For each ReLU neuron \(n_i\), we
write \(n_i^b\) and \(n_i^a\) for its pre- and post-activation values,
respectively, so that \(n_i^a=\max(0,n_i^b)\). This notation separates
the network's linear layers from its nonlinear parts.

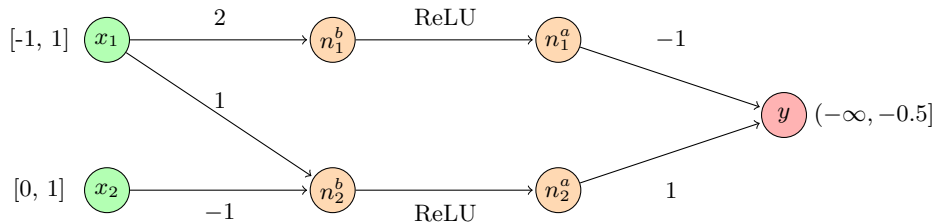
\begin{figure}[htbp]
  \centering
\begin{tikzpicture}[scale=0.9, transform shape, shorten >=1pt, node distance=2cm, auto]
    \tikzstyle{input neuron}  = [circle, draw=black, fill=green!30,  minimum size=17pt, inner sep=0pt]
    \tikzstyle{hidden neuron} = [circle, draw=black, fill=orange!30, minimum size=17pt, inner sep=0pt]
    \tikzstyle{output neuron} = [circle, draw=black, fill=red!30,    minimum size=17pt, inner sep=0pt]
    
    \node[input neuron]  (I-1) at (0,  1) {$x_1$};
    \node at (-0.9,  1) {[-1, 1]};
    \node[input neuron]  (I-2) at (0, 0) {$x_2$};
    \node at (-0.9, 0) {[0, 1]};
    
    \node[hidden neuron] (H-1) at (3,  1) {$n_1^b$};
    \node[hidden neuron] (H-2) at (3,  0) {$n_2^b$};
    \node[hidden neuron] (H-3) at (6,  1) {$n_1^a$};
    \node[hidden neuron] (H-4) at (6,  0) {$n_2^a$};
    \node[output neuron] (O)   at (9,  0.5) {$y$};
    \node at (10.2, 0.5) {$(-\infty,-0.5]$};
    
    \draw[->] (I-1) -- (H-1);         
    \draw[->] (I-1) -- (H-2);          
    \draw[->] (I-2) -- (H-2);          
    
    \draw[->] (H-1) -- (H-3);
    \draw[->] (H-2) -- (H-4);
    
    \draw[->] (H-3) -- (O);           
    \draw[->] (H-4) -- (O);          
    
    \node at (4.5,  1.3) {ReLU};
    \node at (4.5, -0.3) {ReLU};
    \node at (1.5,  1.3) {2};         
    \node at (1.5, 0.7) {1};        
    \node at (1.5, -0.3) {$-1$};       
    \node at (7.5,  1) {$-1$};      
    \node at (7.5, 0) {1};         
\end{tikzpicture}
  \caption{A toy neural network with two inputs.}\label{fig:verification-problem}

\end{figure}

\mysubsection{Running example.}
Consider the network in Fig.~\ref{fig:verification-problem}, with
inputs \(x_1\in[-1,1]\) and \(x_2\in[0,1]\), two hidden ReLU units
\(n_1,n_2\) with pre-activations \(n_1^b,n_2^b\) and post-activations
\(n_1^a,n_2^a\), and a single output \(y\). All bias
values are zero.

For this network, \(n_1^b=2x_1\), \(n_2^b=x_1-x_2\),
\(n_1^a=\mathrm{ReLU}(n_1^b)\), \(n_2^a=\mathrm{ReLU}(n_2^b)\), and
$f(x_1,x_2)=-\,n_1^a+n_2^a \;=\; -\mathrm{ReLU}(2x_1) + \mathrm{ReLU}(x_1-x_2).$
For input \((x_1,x_2)=(0.3,0.7)\), the neuron values are:
$n_1^a=\mathrm{ReLU}(0.6)=0.6,\quad n_2^a=\mathrm{ReLU}(-0.4)=0,$ and $y=-0.6\le -0.5.$

\mysubsection{Neural Network Verification.} We formulate DNN verification as a satisfiability problem.
Let $f:\mathbb{R}^n\to\mathbb{R}^m$ be a neural network,
$X\subseteq\mathbb{R}^n$ an input domain, and $\phi$ a property
over outputs that encodes an \emph{undesirable} behavior (e.g., a
violation of a safety constraint). The verification problem is to
determine whether there exists $x\in X$ such that
$\phi\!\big(f(x)\big)$ holds.  If so, the network
exhibits the undesirable behavior, and the query
is said to be satisfiable (\textsc{SAT}); otherwise, the
network is safe with respect to $\phi$, and the query is 
unsatisfiable (\textsc{UNSAT})~\cite{CaKoDaKoKaAmRe22}.

\sloppy
\mysubsection{Example 1.}
Consider the network in Fig.~\ref{fig:verification-problem}, and
let $X \coloneq (x_1,x_2)\in[-1,1]\times[0,1]$.
Consider the property 
$\phi\big(f(x_1,x_2)\big)\;\equiv\;\big(f(x_1,x_2)\le -0.5\big).$
The verification problem is to determine whether 
$\exists\,(x_1,x_2)\in[-1,1]\times[0,1]:\ f(x_1,x_2)\le -0.5$
holds.
For the specific input \((x_1,x_2) = (0.3,0.7)\), 
the network evaluates to \(f(0.3,0.7) = -0.6 \le -0.5\). 
Thus, \((0.3,0.7)\) is a witness to satisfiability, and the query is \textsc{SAT}.

\subsection{Verification via Branch-and-Bound (BaB)}
DNNs with ReLU activations mix easy-to-solve, linear constraints (affine
layers) with hard-to-solve non-linearities (ReLUs)~\cite{KaBaDiJuKo17}. 
An \emph{activation pattern} is an assignment of active/inactive status to all ReLUs. 
Once an activation pattern is fixed, the network reduces to a single affine transformation, 
and checking feasibility becomes easily solvable. However, a network with $N$ ReLUs induces up to $2^N$ activation
patterns, rendering exhaustive enumeration infeasible.

Many verifiers adopt a branch-and-bound (BaB) search to
tackle the verification problem~\cite{FeNuJoVe22}. BaB
organizes the problem as a binary \emph{splitting tree}: the root encodes the
original verification problem; each node's children
correspond to a case split of one ReLU into its two phases; and
each leaf corresponds to an activation pattern whose satisfiability has been decided.
The process terminates when a SAT sub-problem is
found (the original problem is also SAT); or when all paths in the
search tree are determined to end in an UNSAT node.

In order to prune the search tree, BaB-based solvers usually apply
\emph{bound tightening}: they compute sound bounds on pre-activation
values of the various ReLUs, and use them to identify ReLUs that are
always active or always inactive for the current
sub-problem. Whenever such a ReLU is discovered, there is no need to
branch on it, and this expedites verification significantly.

Another important aspect in BaB-based techniques is the
\emph{prioritization of splits}: choosing which ReLU to split on,
using the information the verifier has obtained so far on the current
sub-problem (e.g., the currently known bounds). Effective
prioritization directs the search quickly toward proofs or
counter-examples, reducing the explored portion of the tree.

\mysubsection{Running example.}
We return to the toy network from Fig.~\ref{fig:verification-problem}
and the query
\[
\exists (x_1,x_2)\in[-1,1]\times[0,1] : f(x_1,x_2)\le -0.5.
\]
Suppose the verifier first branches on \(n_1\), considering the case
\(n_1^b \le 0\). Since \(n_1^b = 2x_1\), this implies \(x_1 \le 0\).
Because \(x_2 \in [0,1]\), it follows that
\(
n_2^b = x_1 - x_2 \le 0,
\)
and so \(n_2\) must also be inactive. Therefore
\(n_1^a = n_2^a = 0\), which yields \(y = 0\). This contradicts the
property \(y \le -0.5\), and thus the entire branch \(n_1^b \le 0\) is
pruned without any further splitting.

Now, consider the case where the verifier first branches on \(n_2\). In the branch \(n_2^b \le 0\),
we obtain only the relation \(x_1 \le x_2\); in the branch
\(n_2^b \ge 0\), we obtain only \(x_1 \ge x_2\). In either case, the
phase of \(n_1\) is not determined. Thus, unlike in the previous case,
the verifier must continue branching.
This illustrates how a good branching decision can lead to a smaller
search space.

\subsection{Splitting Heuristics}
A good heuristic prioritizes splits that are likely either to quickly lead to UNSAT by pruning large, 
infeasible subtrees, or to uncover a satisfying assignment early.
Modern solvers employ several such heuristics, including Sum of Infeasibilities (SoI),
which prioritizes nodes whose assignments in the current, relaxed
solution significantly violate the corresponding ReLU
constraint~\cite{WuZeKaBa22};
the Pseudo-Impact heuristic, which tracks the historical impact of branching on each ReLU~\cite{WuZeKaBa22};
Polarity, which favors ReLUs whose pre-activation bound interval is more balanced around zero~\cite{WuOzZeJuIrGoFoKaPaBa20};
and BaBSR, which scores candidate splits by their estimated contribution to bound
tightening~\cite{BuLuTuToKoKu20}.

\subsection{Reinforcement Learning}
Reinforcement Learning (RL) is a method for producing an autonomous
agent that can make sequential decisions and interact with an
environment, with the goal of maximizing a cumulative reward. The core
components of an RL system are the agent, the environment, states,
actions, and rewards\cite{SuBa18}.
In value-based RL, the objective is to learn an action-value function that estimates 
the long-term utility of taking a given action in a given state.

Double DQN~\cite{HaGuSi16} is a value-based method that improves training stability 
by decoupling action selection from action evaluation through a target network. 
Training typically follows an \(\varepsilon\)-greedy exploration strategy, whereby the agent 
selects the action with the highest estimated value with probability \(1-\varepsilon\), 
and explores alternative actions with probability \(\varepsilon\).

Deep Q-learning from Demonstrations (DQfD)~\cite{HeVePiLaScPiHoQuSeOsDuAgLeGr18} augments Q-learning with expert trajectories. 
The objective combines a value-learning term with a large-margin 
supervised term that encourages imitation of expert actions, enabling training to 
start from demonstration data rather than pure exploration.

\section{Methodology}\label{sec:methodology}
\subsection{Adaptive Splitting Heuristic}
While existing splitting heuristics can significantly reduce the
complexity of DNN verification queries, they operate under a
fundamental limitation: they are static policies, in the sense that a
single heuristic is typically chosen at the beginning of the
verification procedure, and is then applied throughout the search
process.  The performance of these heuristics is highly sensitive to
the specific network architecture and the property being
verified. However, there is often no a-priori way of knowing which
heuristic will be most effective for a given task, and a choice that
works well in one scenario may fail miserably in another.

We argue that it is preferable to consider a splitting strategy that
is not static, but which can instead adapt as the search progresses,
choosing nodes and split variables based on the current state. Many
practical settings involve repeated queries on the \emph{same} network
(often over related input regions or properties): e.g., a
lane-detection model verified under different
conditions~\cite{JuKoOw19,ElElIsDuGaPoBoCoKa24,KeKoCaViFlOtMaMc25}.
Treating each query in isolation discards valuable information about
the network's structure and the efficacy of past branching
decisions. By learning from this cumulative data, we can tap into
information that was previously unavailable.  An adaptive heuristic
can observe the success or failure of past splitting choices and use
this feedback to inform future decisions, effectively transferring
knowledge from one query to the next. 
To achieve these goals, we propose to leverage reinforcement learning
to train an agent that implements a dynamic branching heuristic.

\subsection{Formulation}
\subsubsection{MDP Formulation.}
We frame the ReLU splitting process within a Markov Decision Process (MDP), which allows an RL agent to learn an optimal 
policy for making sequential branching decisions.
Intuitively, the agent's task is to learn a policy that maps the current state to the most effective action, 
aiming to either quickly prune an infeasible branch or find a counter-example.
The MDP in our problem is defined by the tuple $(\mathcal{S}, \mathcal{A}, P, R, \gamma)$:
\begin{itemize}
  \item \textbf{State $\mathcal{S}$}: A state $s\in \mathcal{S}$ represents a specific node in the verification search tree,
  together with a numerical summary of the current subproblem.
  The state is given by a feature vector containing both local and global information. 
  The local features, for each candidate ReLU, include its lower and upper pre-activation bounds, 
  phase status, and heuristic scores such as SoI, polarity, and BaBSR. 
  The global features summarize the search, including the total number of unfixed ReLUs, 
  the current search tree depth, and the number of splits performed so far.
  These features capture information commonly used by 
  splitting-based verification algorithms. 
  \item \textbf{Action} $\mathcal{A}$: An action $a\in \mathcal{A}$ corresponds to 
  selecting an unfixed ReLU and choosing one of its two branching phases
  (\emph{active} or \emph{inactive}). Thus, at each search node, the
  agent scores candidate (ReLU, phase) pairs, and the selected action
  determines the next branching decision.
  \item \textbf{Transition Function} $P(s'|s,a)$: The transition from state $s$ to the 
  next state $s'$ after taking action $a$ is determined by the underlying verifier. This includes 
  bound propagation, relaxation tightening, and constraint feasibility/validity checks. 
  The next state $s'$ is the updated subproblem.
  \item \textbf{Reward (penalty) function} $R(s,a)$: 
  We empirically observed that setting a uniform $-1$ reward per step, which is a common
  choice in RL,  did not guide the search towards
  effective splits well enough ---  presumably due to the large search space.
  We therefore use a task-aligned reward function:
  each split action on neuron $n$ incurs a delayed, normalized penalty proportional 
  to the size of the subtree that action induces.
  Let $k(n)$ be the number of unfixed neurons before action $a$ splits on neuron $n$.
  The potential subtree size, $\mathrm{full}(n) = 2^{k(n)} - 1$,
  upper-bounds the number of internal nodes in a full binary tree with up to $2^{k(n)}$ leaves.
  Let $\mathrm{actual}(n)$ denote the number of internal splits (actions) performed within the 
  subtree rooted at $n$ until that subtree's search is complete (either pruned, or fully explored).
  We assign a single, delayed penalty when the subtree closes:
  \(R(s,a) = -\,\frac{\mathrm{actual}(n)}{\mathrm{full}(n)}.\)
  This normalization makes penalties comparable across neurons with different remaining depth: 
  actions that trigger early pruning yield small (near-zero) penalties, 
  whereas actions that force exhaustive exploration yield penalties near $-1$.
  \item \textbf{Discount Factor} $\gamma\in[0,1]$: Controls the agent's sensitivity to future rewards. 
  A value closer to 1 prioritizes long-term rewards, encouraging the
  agent to consider each split's future impact. We empirically set $\gamma = 0.9$, utilizing this effect
  while avoiding the instability associated with higher discount
  factors.
\end{itemize}

\subsubsection{Leveraging Expert Demonstrations.}\label{sec:Leveraging_Expert_Demonstrations}
Training a DRL agent from scratch proved highly sample-inefficient: 
with sparse effective policies and a complex environment, purely exploratory learning from the available 
state features incurred prohibitively high computational cost and yielded no end-to-end gains. 
To mitigate this, we leveraged existing splitting heuristics as \emph{expert demonstrations}. 
Although sub-optimal, these heuristics encode useful structure; 
using them to guide the policy via imitation pretraining followed by RL fine-tuning provides a viable path to faster, 
more reliable learning than uninformed exploration.

\subsubsection{Q-Learning Integration.}
We solve the MDP using Double DQN with demonstrations (DQfD)~\cite{HeVePiLaScPiHoQuSeOsDuAgLeGr18}. 
Training proceeds in two phases. 
First, the agent is pre-trained on expert transitions generated by running the verifier with 
the heuristics described in Sec.~\ref{sec:preliminaries} (although additional domain- or verifier-specific heuristics could be used as well). 
This learning-from-demonstrations phase optimizes a combined objective consisting of a Q-learning loss and a margin-based imitation loss, 
allowing the agent to learn from the outcomes of expert actions and to mimic them. 
Second, the agent is fine-tuned by interacting with the environment, while continuing to learn from both 
expert- and self-generated transitions. 
Over time, the weight of the imitation term is gradually reduced, allowing the policy to move beyond 
imitation and discover improved splitting strategies. 
During evaluation, the trained policy is deployed directly within the solver's branching loop, where it replaces existing heuristics.

\subsubsection{Hyperparameters \& Trade-offs.}
Training involves balancing learning from expert demonstrations with 
exploration driven by the $\varepsilon$-greedy strategy.
Greater weight on demonstrations or the imitation loss yields a stronger initial policy, 
but may limit the discovery of improved splitting decisions.
We use a standard $\varepsilon$-decay schedule, starting at 1.0 and multiplying by 0.95 each 
iteration until reaching a minimum of 0.05. 

As in most reinforcement-learning settings, there is also a trade-off between training time and final performance; 
accordingly, all reported experiments use a fixed hyperparameter configuration selected in preliminary experiments 
and applied uniformly throughout the evaluation.

\subsubsection{Initial Splits: Pseudo-Impact.}
The Pseudo-Impact heuristic employs a specialized strategy for the initial splits. 
We found this strategy to be efficient for networks with relatively low input dimensions.
We applied a common initial branching policy across all strategies to ensure a fair comparison and to isolate the contribution of our learned heuristic to the deeper, 
more complex portions of the search tree. For all experiments presented in this paper, we applied the Pseudo-Impact heuristic for the first three 
splits of the search tree (depth $\le3$), with the respective
splitting policies taking over from that point onward.

\section{Results}\label{sec:results}
\subsection{Implementation}
To implement our learned branching policy, we integrated a DQfD agent
into the SMT-based verifier \emph{Marabou}~\cite{KaHuIbJuLaLiShThWuZeDiKoBa19}.  As
fixed heuristic baselines, we used \emph{Polarity},
\emph{Pseudo-Impact}, and \emph{BaBSR} (with the first splits always
performed using Pseudo-Impact, as previously discussed).

We evaluated our implementation on the ACAS-Xu networks for
airborne collision avoidance~\cite{JuKoOw19}, under two verification setups:
\begin{enumerate}[(i)]
  \item \textbf{Safety specifications.} The original specifications 
    used in the ACAS-Xu benchmarks, denoted $\phi_1$-$\phi_4$, across all 45 
    ACAS-Xu networks~\cite{KaBaDiJuKo21}, resulting in a total of 180 queries 
    (4 properties $\times$ 45 networks). 

  \item \textbf{Local robustness.} $\ell_\infty$ local-robustness
    verification queries, generated for randomly selected inputs
    on ACAS-Xu network $N_{1,1}$. For an input $x_0$, a local
    robustness query verifies that the network's predicted advisory
    remains unchanged throughout the ball $\|x-x_0\|_\infty \le \delta$ around
    $x_0$.  We tested 1000 randomly selected inputs with three
    radius values: $\delta \in \{0.08,0.09,0.1\}$, producing a total
    of 3000 queries. 
    The varying inputs and robustness radii
    produced a mix of satisfiable, unsatisfiable, and timed-out instances,
    allowing for a more informative comparison between strategies.
  \end{enumerate}

  We trained a separate agent for each of the two setups. Training
  included a DQfD warm start followed by Double DQN and used
  approximately \(5\%\) of the queries available in the respective benchmark;
  and evaluation was then performed on the benchmark's remaining queries.

The training budget was fixed a priori to \(45{,}000\) splitting steps: 
\(5\) epochs \(\times\) \(1{,}000\) steps with demonstrations (DQfD), followed by 
\(40\) epochs \(\times\) \(1{,}000\) steps with
self-exploration/exploitation (DDQN).
Each agent required about 2 hours of offline training.
All experiments were conducted on single-CPU machines with 2 GB of memory, running
Debian~12, and with a 1-hour timeout.
 
\subsection{Performance}
For each splitting policy, we report the number of queries where a
counter-example was found
(\emph{SATs}), unsatisfiable queries (\emph{UNSATs}), and runs that exceeded the time limit
(\emph{Timeouts}). We also report the average time per instance
(\emph{Avg Time}) and the average number
of solver main-loop iterations (\emph{Avg Iterations}). Both averages are computed over all queries.
For runs that exceeded the time limit, we use the timeout threshold as the runtime.

\subsubsection{Setup (i): ACAS Xu Safety Properties.}
On the original ACAS Xu properties, our learned splitting heuristic
outperforms all other heuristics in the number of solved instances and
average verification time. The aggregate results appear in
Table~\ref{tab:acas_summary}, demonstrating the effectiveness of our
method.  \begin{table}
  \centering
\caption{ACAS-Xu runs across all networks: averages over all instances, with timeouts counted as 1 hour.}\label{tab:acas_summary}
\setlength{\tabcolsep}{4pt}
\small
\begin{tabular}{lrrrrr}
\toprule
Heuristic & SAT & UNSAT & Timeout & Avg.\ Time (ms) & Avg.\ Iter. \\
\midrule
DQfD Agent & \textbf{42.00} & \textbf{105.00} & \textbf{23.00} & \textbf{692691.04} & 838986.76 \\
Polarity & 40.00 & 103.00 & 27.00 & 771287.04 & 795486.79 \\
Pseudo-Impact & 23.00 & 84.00 & 63.00 & 1347021.98 & 1183586.92 \\
BaBSR & 41.00 & 104.00 & 25.00 & 775690.24 & \textbf{779218.49} \\
\bottomrule
\end{tabular}
\end{table}

\subsubsection{Setup (ii): Local-Robustness Properties.}
Table~\ref{tab:robust_summary} summarizes the results for setup (ii).
The learned splitting heuristic achieves the best average verification
time, although it requires a slightly greater number of iterations
compared to the Polarity heuristic. Moreover, it solves a
significantly higher number of instances compared to the
baselines.\begin{table}
  \centering
\caption{Summary of robustness runs for common (ExampleID, Epsilon).}
\label{tab:robust_summary}
\setlength{\tabcolsep}{4pt}
\small
\begin{tabular}{lrrrrr}
\toprule
Heuristic & SAT & UNSAT & Timeout & Avg.\ Time (ms) & Avg.\ Iter. \\
\midrule
DQfD Agent & \textbf{253.00} & \textbf{2595.00} & \textbf{122.00} & \textbf{328457.31} & 235690.92 \\
Polarity & 252.00 & 2574.00 & 144.00 & 348988.84 & \textbf{222903.82} \\
Pseudo-Impact & 88.00 & 2319.00 & 563.00 & 749976.29 & 732681.33 \\
BaBSR & 244.00 & 2550.00 & 176.00 & 376507.73 & 265502.52 \\
\bottomrule
\end{tabular}
\end{table}

\begin{figure}[t]
\begin{subfigure}{0.48\textwidth}
    \centering
    \includegraphics[width=\linewidth,trim=0.3cm 0.3cm 0.3cm 0.3cm,clip,alt={A line chart titled ``Performance (ACAS-Xu)'' showing the ``Number of Instances Solved'' on the y-axis against ``Time (seconds)'' on a logarithmic x-axis. It compares four methods: DQfD Agent, Polarity, Pseudo-Impact, and BaBSR. All methods show a steep initial increase, briefly plateauing near 80 instances solved between 1 and 10 seconds. After 10 seconds, the Pseudo-Impact method remains mostly flat, ending near 90 instances. The other three methods resume climbing, reaching roughly 130 instances solved by the end of the time limit, with the DQfD Agent solving slightly more instances than Polarity and BaBSR.}]{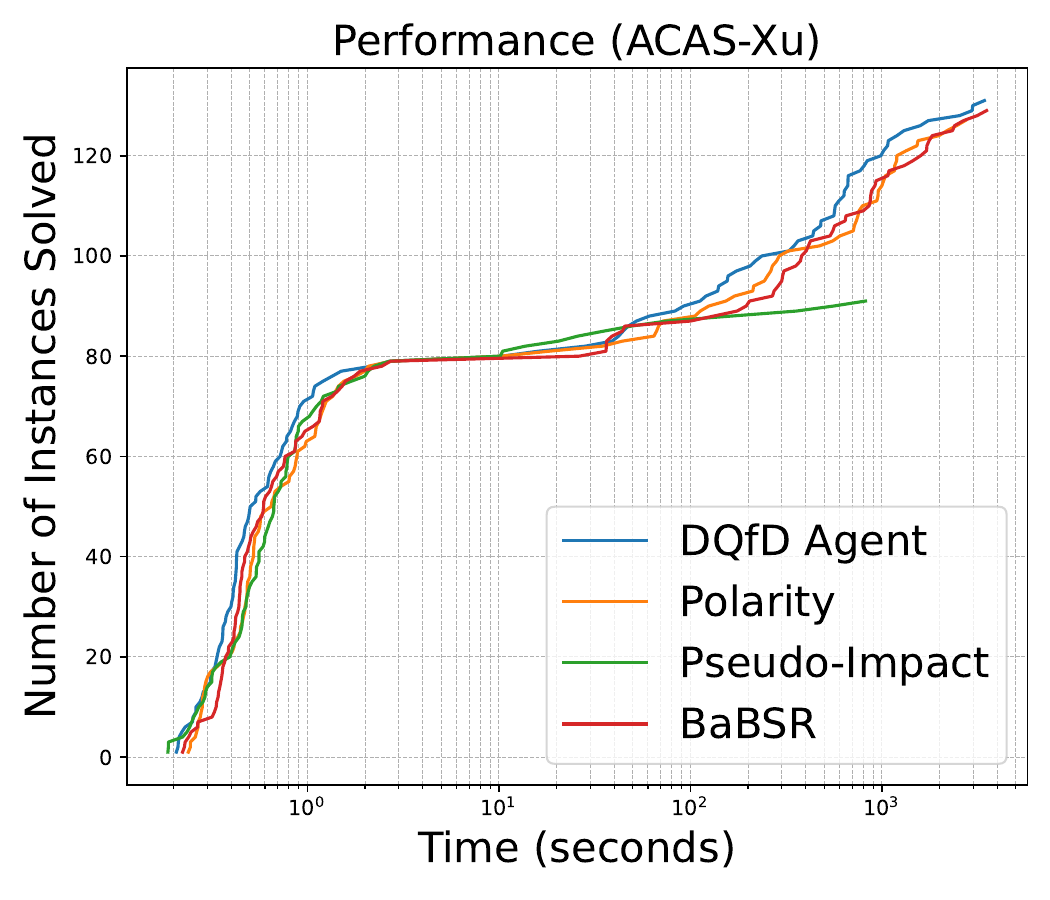}
    \caption{Setup (i).}\label{fig:acas_log}
  \end{subfigure}\hfill
  \begin{subfigure}{0.48\textwidth}
    \centering
    \includegraphics[width=\linewidth,trim=0.3cm 0.3cm 0.3cm 0.3cm,clip,alt={A line chart titled ``Performance (ACAS-Xu Robustness)'' showing the ``Number of Instances Solved'' on the y-axis against ``Time (seconds)'' on a logarithmic x-axis. It compares the same four methods: DQfD Agent, Polarity, Pseudo-Impact, and BaBSR. The methods initially climb steeply together. After roughly 1 second and 1000 instances solved, the Pseudo-Impact method's rate of solving slows down considerably, finishing lowest at roughly 1600 instances solved. The DQfD Agent, Polarity, and BaBSR methods remain tightly clustered together, steadily increasing to finish just above 2000 instances solved.}]{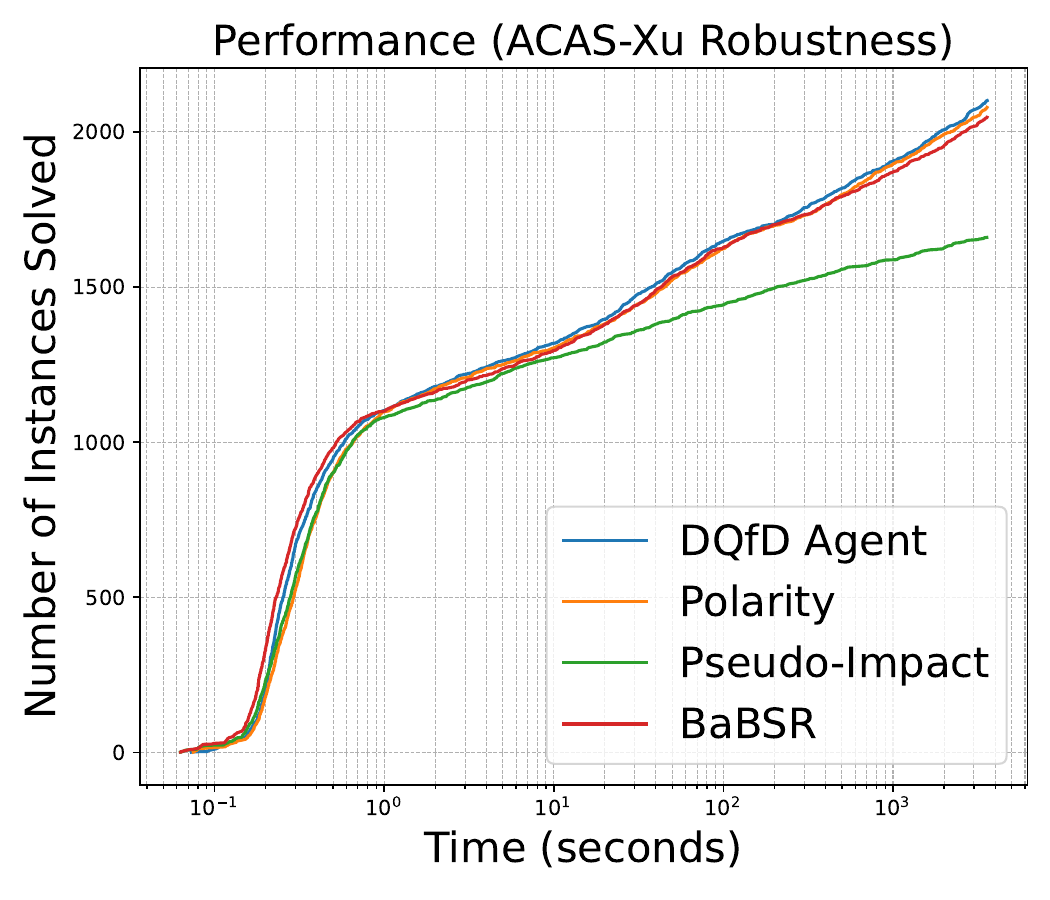}
    \caption{Setup (ii).}\label{fig:robustness_log}
  \end{subfigure}
  \caption{Cumulative instances solved vs.\ time (log scale) on the
    two setups.}
  \label{fig:cumulative_times}
\end{figure}

Fig.~\ref{fig:cumulative_times} demonstrates
that the learned splitting heuristic closely tracks the strongest
baseline on easier instances --- indicating an effective imitation of
expert behavior. On harder instances, however, it surpasses the baseline, suggesting
more effective splitting choices.

\section{Related Work}\label{sec:related_work}
\mysubsection{Exact verification via SMT and BaB.}
Complete solvers repeatedly perform case splitting on piecewise-linear activations and propagate bounds to 
prune the search space~\cite{KaBaDiJuKo17,KaHuIbJuLaLiShThWuZeDiKoBa19}. 
Recent work strengthens this paradigm through tighter relaxations and improved branching scores~\cite{BuLuTuToKoKu20}. 
Our contribution is complementary: we learn a splitting policy that
can be inserted into a BaB framework or an SMT-based
solver, replacing hand-crafted splitting heuristics.

\mysubsection{Learning for branching and verification.}
Learning-guided branching has been explored in several related settings.  
In neural network verification, Jeong et al.~\cite{JeLuKe21} use a graph neural network to guide \emph{node selection}. 
Our work is distinct: we use DQfD to learn the \emph{branching rule} itself, directly affecting how subproblems are created and, 
consequently, the effectiveness of bound tightening and pruning. 
Qu et al.~\cite{QuLiZhZeYuWaLvLiMa22} apply reinforcement learning to
branching in general MILP queries; 
however, DNN verification is structurally different, because branching must interact with specialized 
nonlinear relaxations and sound bound-tightening procedures. Liang et al.~\cite{LiGaPoCz16} propose an online 
adaptive branching heuristic for CDCL SAT solvers; while similarly motivated, their setting involves Boolean branching with 
online feedback, whereas our policy is learned offline from demonstrations and transferred across related verification queries.

\section{Conclusion and Future Work}\label{sec:conclusion_and_future_work}
We introduced a novel reinforcement-learning-guided splitting
heuristic for DNN verifiers. Our method integrates learning from
demonstrations with Double DQN inside an SMT-based
verifier, and improves verification performance speed.

Several directions remain for future work, aimed at making learned
branching a more reusable and scalable.
First, the current framework can be improved by enriching the MDP
representation. Additional state features and alternative reward
signals may better inform the agent's decisions and help optimize
overall verification performance.
Second, our evaluation focuses on ACAS Xu and feed-forward networks
with ReLU activations. Assessing the broader applicability of the
approach will require experiments on a wider range of benchmarks and
network architectures.
Finally, an important goal is to improve the portability of learned
splitting policies across networks and properties. This includes
studying transfer across networks with different input sizes and
architectural variations. Progress in this direction could
significantly reduce the cost of applying learned branching to new
verification tasks.

\begin{credits}
  \subsubsection{\ackname} 
This work was partially funded by the European Union
(ERC, VeriDeL, 101112713). Views and opinions expressed
are however those of the author(s) only and do not necessarily reflect those of the European Union or the European
Research Council Executive Agency. Neither the European
Union nor the granting authority can be held responsible for
them. This research was additionally supported by a grant
from the Israeli Science Foundation (grant number 558/24).

\subsubsection{\discintname}
The authors have no competing interests to declare that are
relevant to the content of this article. 
\end{credits}


\section{Data-Availability Statement}
The code and experiments reported in this paper are available online~\cite{SwKa25}.

\bibliographystyle{abbrv}
\bibliography{RL_Splitting}

\end{document}